# Monitorology – the Art of Observing the World

Miroslaw Malek

Advanced Learning and Research Institute (ALaRI)

Faculty of Informatics

USI- Lugano, Switzerland

**Abstract** – In the age of ever-increasing demand for big data and data analytics, a question of collecting the data becomes fundamental. What and how to collect the data is essential as it has direct impact on decision making, system operation and control. Specifically, we focus on the art of observing the world by electronic devices such as sensors and meters that, in general, we call monitors. We define five challenges to ensure effective and efficient monitoring that still need a lot of research. Additionally, we illustrate each challenge by example.

Since reliance on big data and data analytics is continuously increasing, these challenges will become ever more relevant to save the world from flood of meaningless, dumb data, leading frequently to false conclusions and wrong decisions whose impact may range from a minor inconvenience to major disasters and even loss of lives.

**Where is the wisdom we have lost in knowledge? Where is the knowledge we have lost in information?**

**Thomas Stearns Eliot (1888-1965)**

**Where is the information we have lost in data?**

1. Introduction

Observation, monitoring or data acquisition, followed by data collection, has been the most common and perhaps the most successful scientific method since the beginning of times. It is not only used in science to observe the world and state hypothesis but it has also been used virtually in all domains and all walks of life from archeology and business to physics and zoology. Formally, monitoring is observation and collection of relevant data about the current state of a system under study. The purpose of monitoring may vary from noble causes such as better understanding of the world and saving lives, decision making and efficient management to dictatorship, blackmail and espionage not excluding impinging on privacy. Monitoring is also used to ease or optimize control of machines, including robots and vehicles, or to diagnose a system status. In electronic and mechanical systems (hardware) typically physical features such as temperature, load or pressure are measured by sensors while in software log files or probes are used. Machine learning, predictive analytics and artificial intelligence are all based on the vast amount of collected data. In this article, we focus mainly on technical aspects of monitoring but the ethical issues are equally important and diverse, therefore requiring a separate treatment.

In the flood of data generated daily, it is not easy to filter out the relevant information, but even more challenging is to infer the knowledge, not mentioning the wisdom that even teams of experts are not able to derive in majority of real-life situations (see Figure 1). Although several classification methods ranging from statistics and machine learning to pattern recognition and data mining exist, knowing what to collect regarding data or information might be more effective than a particular method itself.

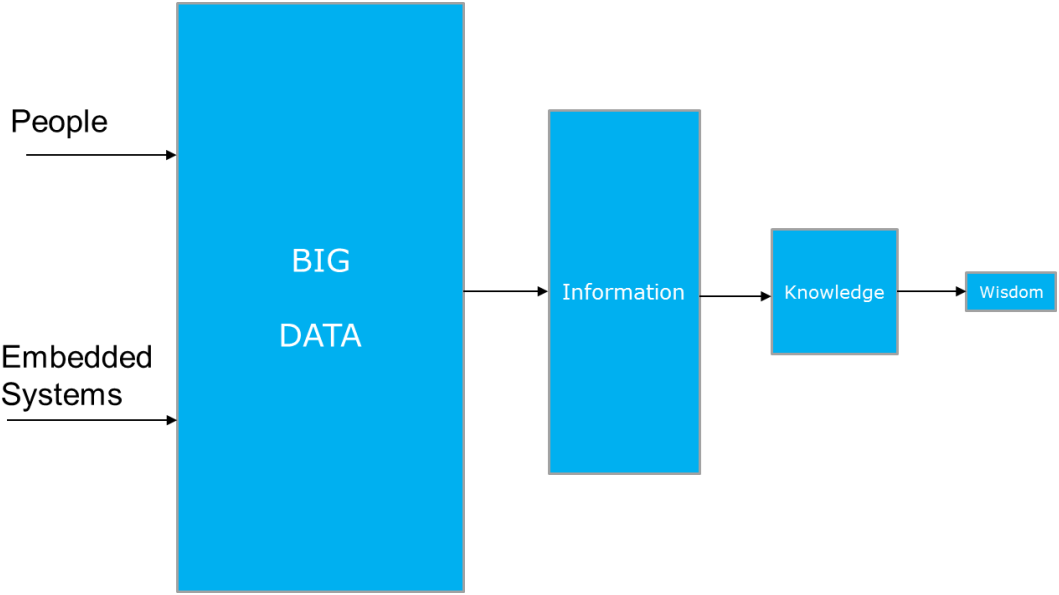

Figure 1: Challenge of getting Small Data out of Big Data in a form of information, knowledge or wisdom.

The biggest generators of the masses of data are we, the humans, along with cyber-physical and embedded systems which are monitoring both the nature (climate, environment, including humans themselves, etc.) and artificial world created by us. These include industrial processes, means of transportation and communication, software, organizations, our domiciles, factories, offices and practically everything else.

With incredible progress in embedded systems ranging from smart meters of all kinds to smartphones, we observe an explosive growth of generated data (so called "Big Data") and the fundamental challenge is how to make the Big Data small and extract meaningful answers to posed questions, simplify it, make right decisions or check validity of hypotheses. In other words, the question is how to distill out of vast amounts of raw, dumb data, the information, knowledge or wisdom. One of the keys to meaningful observations is to determine first what features (also called variables, parameters or events in different research communities) of a system or a phenomenon are most relevant or most indicative. In order to find out, a number of preparatory steps has to be taken. To reach this stage methodically, all available data should be collected first, followed by the process of feature (variable) selection in order to identify the most indicative variables as well as invariants and correlations for a given purpose.

We need to keep in mind that a permanent strive for more data (data for the sake of data), if improperly performed, may result in loss of the information content that we are looking for, or even worse, lead to misinformation and wrong decisions.

As acutely observed in [3], businesses and governments exploit big data without regard for issues of legality, data quality, disparate data meanings, and process quality. This often leads to wrong or poor decisions where institutions and/or individuals are affected. Therefore, it is yet another argument for knowing what, when and how to collect the data.

Despite a large number of papers on monitoring, it still remains more of an art than a science. This article attempts to list major challenges and shed some light on how they might be tackled.

Furthermore, the data is the main asset of most corporations, governments, institutions and individuals. Its quality, scope and size will have ever bigger impact on the way we think, learn, live, work, produce and create. The issue at hand is, in fact, bigger than the Big Data, as it may influence the decision making process in all walks of life including politics, economics, technology and society.

## 2. Monitoring Objectives

People, animals and machines observe/monitor the world with a variety of objectives that relate to the past, present or future. Inspired by Gartner's vision [5], we divide objectives into four categories with respect to time where the data is collected in order to analyze the past (what happened?), diagnose the present (why did it happen?), predict the future (what will happen?) or construct the future (how can we make it happen?). In short, the ultimate goals are to understand the past and/or observe and/or control the present or control the future. The level of difficulty increases as we move along the time axis and the potential value goes up as we move from analysis and diagnosis to prediction and constructing the future. Both analysis and diagnosis can be considered as reactive methods/algorithms that are applied upon an occurrence of an event while methods and algorithms that predict and construct the future belong to a category of proactive approaches. These distinctions are summarized in Table 1.

| Past | Present | Future | |
|---|---|---|---|
| What happened? | Why did it happen? | What will happen? | How can we make it happen? |
| Analyze | Diagnose | Predict | Construct |
| Reactive | | Proactive | |

**Table 1** Monitoring Objectives with respect to Past, Present and Future.

Since observation/monitoring is the first step to almost any activity, not surprisingly it plays a pivotal role in decision making, cooperation, management, diagnosis, adaptation in many other diverse processes and activities.

Consequently, it is evident that monitoring plays a key role in most activities in nature and in the artificial world created by humans. Since monitoring is so essential, in the next section we identify the main challenges in monitoring optimization.

### 3. Main Challenges

We now propose and examine five main challenges concerning monitoring, describe current approaches, provide guidelines for the future and illustrate each challenge by example in failure prediction and malware detection.

Challenge 1: **What to monitor?**

This question depends on our application and its goal function. In today's computer we can monitor tens of thousands of its variables, probably millions in a human, but in reality we monitor much fewer of them as we are usually interested in very specific properties such as performance, reliability, security or timeliness or, in the case of a human, it might be a specific health condition or a habit.

We are also limited by technology and the knowledge of the processes or products. We still do not have good monitors for wine but milk that is just about to turn sour, can be identified indirectly by measuring its lifetime and temperature and understanding its bacteria growth process. If we have more exotic goals such as failure prediction, emotional state or privacy protection, we may need to actually use some feature selection algorithms in order to find out what features are the most significant ones.

A number of approaches to feature selection is well summarized in [7] while a comprehensive survey on feature selection algorithms for classification and clustering can be found in the paper by Liu and Yu [8]. In our failure prediction methodology [9], [10] we found out that feature selection has bigger impact on precision (the ratio of the number of true-positive alarms to the total number of alarms) and recall (the ratio of the number of true-positive alarms to the total number of failures) than the choice of the model.

We have examined over 1100 features using regression analysis methods and with over 110 million experiments we were able to select two features (the number of exceptions per second and the growth rate of the operating system memory), that were most indicative regarding failure prediction in telephone servers, which resulted in 0.83 precision and over 0.8 recall [9].

In other examples, what to monitor might be regulated by public authorities, e.g., air, water or food quality, by a given enterprise for the purpose of efficient management or it might be dictated by the manufacturing process and its goals.

Challenge 2: **Where to place the monitors in a system? How many monitors do we need?**

The placement problem has always been a challenge: from placement (layout) of transistors on a chip to placement of nodes in wireless network for maximum connectivity. The same holds for placement of monitors.

Several solutions exist and are usually problem-specific because optimization goals vary: minimization of chip area for placement of transistors and maximum coverage for a given area with minimum number of nodes like sensors or wireless communication chips. For monitor (sensor) placement, the objective is to get the most relevant data at required frequency (sampling rate) at minimum cost. Placement costs may vary significantly if, for example, some monitors have to be placed in space, underground, underwater or on a steep mountain. The complexity of a problem increases when we go from a two-dimensional to three-dimensional placement. The quality of placement such as coverage has direct impact on the minimum number of required sensors.

But the minimum number of monitors does not necessarily mean optimum as the placement may have additional optimization criteria. How will the accuracy of the measurement be affected by larger number of monitors? Of course, cost plays a critical role and in most industrial systems the number of monitors is kept to the minimum unless an additional requirement such as fault tolerance necessitates redundancy.

There is a number of sensor placement problems and solutions [12], [20] and some of them might be adopted for the monitor placement. Specifically, for monitoring of a mobile object or a moving human, the triangulation placement method can be used. It requires that each monitored object is covered by at least three sensors/monitors. The algorithms usually optimize the number of required nodes. Since the node minimization and placement problem is NP-hard [2], we frequently use heuristics that are not only fast but usually provide a good solution.

Additional questions that need to be asked when deciding on the placement and the number of monitors are: how reliable, how secure they are, how much power they use and whether or not they are required to operate in real time.

Reliability of monitors should be assessed in advance as the failure of a monitor may have severe consequences. Is the monitoring infrastructure fault tolerant? Is it able to cope with a failure of one, two or even k monitors?

Security is another key issue that should be addressed a priori because monitor manipulation may have dire consequences. Making sure that monitoring reflects the reality under operating conditions of a system or a device is another aforementioned challenge.

Since monitors are add ons to a system operation, their power requirements must also be assessed a priori and, typically, they should use as little power as possible (consider low power design) and be noninvasive.

Finally, the question of time and real time has to be considered. Typically, multiple monitors need a common time base and a simple GPS-based synchronization might not be sufficient. Therefore, in some cases we need to resort to the Universal Time Coordinated (UTC) using sophisticated synchronization protocols.

Additionally, if monitors must deliver measurements in real time, system designers must ensure meeting deadlines and durations through appropriate scheduling policies and the Worst Case Execution Time (WCET) analysis.

Typical example here is a placement of monitoring sensors that track a person in a factory environment to ensure his/her safety. Given a factory floor with obstacles, modeled by grid, we should place monitoring sensors such that they guarantee full coverage at the minimum cost. Several solutions exist that may result in substantial cost savings in terms of the number of sensors placed on a grid [20].

In some cases, the number and the placement of the monitors can be prescribed by government regulators as in the case of, e.g., air or water pollution monitoring.

Challenge 3: **When or how frequently to monitor?**

The question of how frequently a certain feature should be measured is a fundamental one and ranges from billions of samples per minute to one per day, per month or per year.

Formally, we define the sampling rate, sample rate, monitoring frequency or sampling frequency as the number of samples per second (or per other unit of time or event) taken from a continuous or discrete signal to make a discrete signal of a given frequency. For time-domain signals, the unit for sampling rate is hertz (inverse seconds, $1/s$, $s^{-1}$). For example, a sampling rate for a phone is 8 kHz while High-Definition DVD requires 192 kHz. The inverse of the sampling frequency is the sampling period or sampling (monitoring) interval, which is the time between samples [15], [19]. A wide body of literature exists on this topic for many diverse applications.

There are three main monitoring policies: 1) time triggered; 2) event triggered, and 3) a hybrid. Choosing a monitoring frequency depends on an application and many other factors such as the goal, quality of a result, observation period as well as the effect on performance, storage, communication and processing capabilities. It is frequently a tradeoff specific to a given application between cost and quality of the result or decision making.

Time-triggered monitoring requires a good understanding of a process in order to optimize the sampling frequency. Event-triggered monitoring focuses on observing changes in a system and might be more efficient, especially in stable systems. Typically, in complex systems a hybrid approach is used as it allows to tailor monitoring of each feature according to the needs. In some cases, adaptive monitoring can be used where sampling frequency changes depending on the state of the system. In failure prediction, for example, we may increase monitoring frequency if we observe that the system is in a dangerous state.

In another example, time-triggered monitoring of an electric grid at the frequency higher than 50 Hz might not make sense because electric grid operates at 50 Hz, and therefore, the sampling period does not have to be shorter than 20 ms. The question is whether from an application perspective such monitoring improves grid's stability or not. If it takes us 40 ms to process the data then we might be forced to be satisfied with 25 Hz sampling rate as we might not be able to process the data. On the other hand, if we have two processors, we may interleave them such that we can handle a 50Hz sampling rate even with 40 ms processing time. Ultimately, the sampling frequency depends on the purpose, the hypothesis that is posed or on an application.

Adaptive monitoring, may also be useful in malware detection because we may increase the rate of monitoring when we observe that the system might be in endangered state. In malware detection [14] we have used different sampling intervals (inverse of sampling rates) of 2, 4, 6, 8, 12 and 16 seconds and have found out that if the alarm triggering threshold is appropriately adjusted, the quality of detection decreases only slightly with increase of monitoring interval. One important lesson from our experiments: sampling frequency may depend on many different factors so optimizing it might be a highly-complex problem.

Another example for choosing a sampling rate in an IoT device also heavily depends on design goals. If it is accuracy, try to sample as fast as possible. If it is minimizing storage, processing and communication time, sample as slow as possible. If your goal is to get a certain response time, find out what is the optimum rate to reach this goal. Sometimes the sampling rate is determined by the execution time of the decision algorithm or physical limits of the device or environment. Pay attention to data quality, as factors such as noise or instability may influence the preferred sampling rate as well.

In addition, we need to decide for how long we must observe a given phenomenon to draw some credible conclusions in order to make control or management decisions. So together with a sampling rate, we should know at what speed or rate the output/decisions need to be made to ensure proper functioning of a system or organization. Furthermore, the best threshold value for classifiers should be determined such that the F-measure is optimized ensuring high precision and recall. There is no silver bullet here and, in many cases these values are obtained experimentally as the diversity of applications is immense.

Like in the previous challenge, the monitoring frequency may also be prescribed by the government regulators, e.g., air or water pollution sampling frequency.

Challenge 4: **What, where, when and how to communicate, store and process the monitoring data?**

Depending on the goal, efficiency and strategy (centralized, distributed or hybrid monitoring) the communication may require a significant bandwidth to transmit the monitoring data, therefore, the questions like what, where, when and how to communicate have to be addressed a priori.

In centralized monitoring, all monitors send the data to a single computer that is in the position of not only observing the status of each monitored device but also identifying trends for the entire groups of devices or monitors.

In distributed monitoring each device is monitored autonomously and it is used in the cases when communication is impossible or too expensive.

The third option is a hybrid where some values are observed locally, some might be even partially processed and then the rest of them are sent continuously or periodically or periodically in batches to a central computer. Which mode of operation to choose strictly depends on an application and user requirements.

Monitoring may produce an immense amount of data. What about how, where and when to store such data? If the data comes from multiple sources (e.g. sensors), then it is typically unstructured and arrives at different intervals. The question how this data can be stored should address the format, database organization, synchronization and storage devices. It is important to create comprehensive and expressive representation of collected data in a form of, for example, log-files that enable a flexible and semantically augmented representation of the logged events which in turn can be analyzed automatically [17]. The next question is whether the data should be stored locally, next to monitored system or monitoring device or centrally to enable comparative analysis or a more general system view? The classical database systems are geared towards static accumulation of vast amount of data whose storage is mainly controlled by humans while monitoring systems generate data with varying frequency and/or sporadically in case of event monitoring. Furthermore, usually the latest version of the data is easily accessible while earlier logs are archived. This is usually not acceptable in, for example, machine learning applications where typically the entire sequences of monitoring data are required. An example of a database, addressing most of these problems, is Aurora database [1].

Finally, we should decide where, when and how to process monitoring data. Deciding whether to process the data locally or centrally will have a direct impact on processing and communication time. It may turn out that a hybrid approach is most efficient where data is processed locally and only the relevant outcomes are passed on to a central host. Another hybrid could be that some data is processed locally and the remaining data centrally. This depends on, for example, the need for local and for global information or on processing overhead where some parts of application are processed locally and the rest is offloaded to a central server or a cloud [4]. Again and again, the methods and timing of processing the data have to be adapted to the posed questions, applications and goals.

In many applications, including our running example of failure prediction, the data is stored locally and processed on the same server. For more complex prediction algorithms a coprocessor or a graphics processing unit (GPU) can be used.

Here also, what, where, when and how to communicate, store and process the monitoring data, in some cases (e.g. environmental protection), may be prescribed by the government regulators, considering additional requirements concerning data privacy and security.

Challenge 5: **How good is the quality of data that we get?**

Data quality is the reliance that users can put on the acquired data in terms of precision and accuracy in order to obtain a faithful reflection of monitored world. Once collected the monitoring data should remain unchanged (data stability). In the nutshell, an ideal high-quality data should be complete, adhere to standards, consistent, stable over time, accurate and time stamped. According to [6] data quality can be characterized by four attributes: accuracy, availability, interpretability and timeliness. The problem is that accuracy has many definitions. The accuracy of a measurement system is the degree of closeness of measurements of a quantity to that quantity's actual (true) value [6]. The precision of a measurement system is related to reproducibility and repeatability. Additional characteristics of data may refer to completeness, severity of inconsistency (anomalies) and missing or unknown variables.

Data quality has been researched by many and good surveys can be found in [18] and more recently in [13] and [16]. The layer of software which helps to measure and collect the data can be manipulated and can range from an obvious deception as in the VW emission affair [11] to small inconsistencies that can produce a completely different picture of reality over time. The

VW case shows that public authorities and companies have to pay more attention to how monitoring is conducted, under what conditions and what procedures to incorporate to ensure the true values of the measured entities. These problems require a serious consideration because what we have seen so far is just a tip of the iceberg. So one of the main challenges is indeed data quality assurance. Additional problem may be caused by monitoring devices which may fail or may skew the outcome due to interference and impact on measurements or inherent imprecision.

## 4. Monitoring Guidelines

Research, design and development of any technical system or creation and management of any organization, be it government or an industrial enterprise, should start with quantifiable goals. Once we know what we want to accomplish we can define measurable, quantifiable variables or features that are often translated into key performance indicators (KPI).

Monitoring goals and methods must be a part of system specification. As indicated in Challenge 1, what to monitor is fundamental. This question is often answered based on gut feeling or previous experience while scientific methods should be used [7], [8].

Once we know what to monitor, we need to answer the question how: do we have sensors to monitor variables that will result in optimized decision making or give us assessment of KPIs? How many monitors will we need and where they should be placed (Challenge 2).

With answers comes the next question: what should be the sampling frequency or do we need a continuous observation of a given phenomenon to control or understand it (Challenge 3)?

Once we find solutions to these three aspects of system design or enterprise management, we need to find the means of storing, transmitting and processing the acquired data (Challenge 4). In this phase we also should take care of the issues related to data privacy and security.

Once we have monitoring system in place, we can test it, assess its quality, accuracy and its precision (Challenge 5). In the process of system design, implementation and exploitation the monitoring should be continuously refined to deliver highest quality data at low cost to facilitate optimized control, decision making and management.

The illustration of the iterative monitoring process is shown as a cycle in Figure 2.

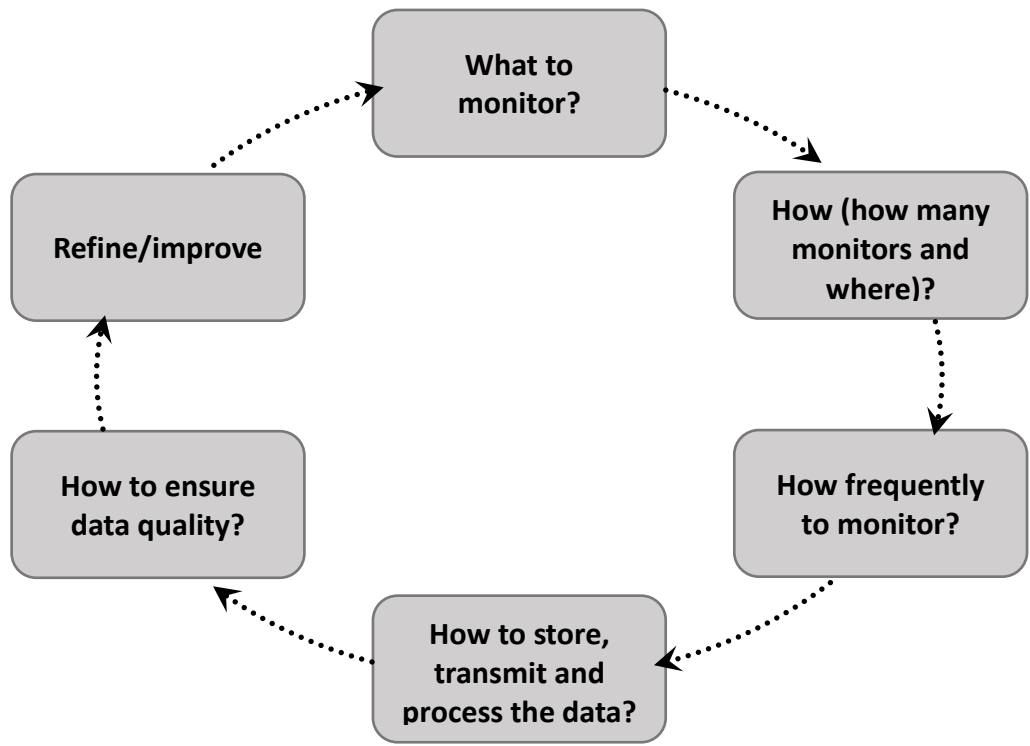

Figure 2. A cycle of iterative monitoring process based on five challenging questions.

## 5. Conclusions

With ever-growing hunger for data and unbounded potential of data analytics, we need to focus more on the front end of the process, namely, data acquisition. This in turn requires more research on what we propose to call monitorology or the art of observing the world. Despite a lot of research and experience in automated data acquisition, several, fundamental issues remain open and methodologies tailoring a data acquisition system to any application still need to be refined.

Five challenges were posed and addressing them will make the process of distilling information from data, acquiring knowledge from information and sometimes inferring wisdom from knowledge more accurate, precise, more complete and useful. This will have in turn a significant impact on improving quality of decision making, acquiring deeper knowledge and ultimately building a better world around us.

**Miroslaw Malek** is Professor at Advanced Learning and Research Institute (ALaRI) at the Faculty of Informatics at the University of Lugano (USI) that specializes in properties of cyber-physical and embedded systems. Prior to that he spent 17 years at the University of Texas at Austin (1977 - 1994) where he was also a holder of the Bettie Margaret Smith and the Southwestern Bell Professorships in Engineering. From 1994 until 2012 he was professor and holder of Chair in Computer Architecture and Communication at the Department of Computer Science at the Humboldt University in Berlin. His research interests focus on dependability, security and real time in diverse computing environments ranging from clouds to Internet of Things.